\documentclass[twocolumn,showpacs,aps,prl,superscriptaddress,floatfix,amsmath,amssymb]{revtex4}

\usepackage{graphicx}
\usepackage{dcolumn}
\usepackage{bm}
\usepackage{subfigure}

\begin{document}

\title{Antiferromagnetic transitions in `$T$-like' $BiFeO_{3}$}

\author{G. J. MacDougall}
\affiliation{Quantum Condensed Matter Division, Oak Ridge National Laboratory, Oak Ridge, Tennessee, USA, 37831}

\author{H. M. Christen}
\email{christenhm@ornl.gov}
\affiliation{Materials Science and Technology Division, Oak Ridge National Laboratory, Oak Ridge, TN, USA, 37831}

\author{W. Siemons}
\affiliation{Materials Science and Technology Division, Oak Ridge National Laboratory, Oak Ridge, TN, USA, 37831}

\author{M. D. Biegalski}
\affiliation{Center for Nanophase Materials Sciences, Oak Ridge National Laboratory, Oak Ridge, TN, USA, 37831}

\author{J. L. Zarestky}
\affiliation{Department of Physics and Astronomy, Iowa State University, Ames, Iowa 50011, USA}
\affiliation{Division of Materials Science and Engineering, Ames Laboratory, Iowa State University, Ames, Iowa 50011, USA}

\author{S. Liang}
\affiliation{Materials Science and Technology Division, Oak Ridge National Laboratory, Oak Ridge, TN, USA, 37831}
\affiliation{Department of Physics and Astronomy, University of Tennessee, Knoxville, TN, USA, 37996}

\author{E. Dagotto}
\affiliation{Materials Science and Technology Division, Oak Ridge National Laboratory, Oak Ridge, TN, USA, 37831}
\affiliation{Department of Physics and Astronomy, University of Tennessee, Knoxville, TN, USA, 37996}

\author{S. E. Nagler}
\affiliation{Quantum Condensed Matter Division, Oak Ridge National Laboratory, Oak Ridge, Tennessee, USA, 37831}
\affiliation{CIRE, University of Tennessee, Knoxville, TN, USA, 37996}

\date{\today}
\begin{abstract}

Recent studies have reported the existence of an epitaxially-stabilized tetragonal-like (`$T$-like') monoclinic phase in $\mathrm{BiFeO_{3}}$ thin-films with high levels of compressive strain. While their structural and ferroelectric properties are different than those of rhombohedral-like (`$R$-like') films with lower levels of strain, little information exists on magnetic properties. Here, we report a detailed neutron scattering study of a nearly phase-pure film of $T$-like $\mathrm{BiFeO_{3}}$. By tracking the temperature dependence and relative intensity of several superstructure peaks in the reciprocal lattice cell, we confirm antiferromagnetism with largely G-type character and $T_{N}$ = 324 K, significantly below a structural phase transition at 375 K, contrary to previous reports. Evidence for a second transition, possibly a minority magnetic phase with C-type character is also reported with $T_{N}$ = 260 K. The co-existence of the two magnetic phases in $T$-like $\mathrm{BiFeO_{3}}$ and the difference in ordering temperatures between $R$-like and $T$-like systems is explained through simple Fe-O-Fe bond distance considerations.
\end{abstract}


\maketitle

The oxide-perovskite $\mathrm{BiFeO_{3}}$ is the only known room temperature multiferroic\cite{catalan09}, with robust ferroelectricity below $T_{c}$ = 1103 K and long-wavelength cycloidal antiferromagnetism below $T_{N}$ = 643 K in bulk form\cite{ramazanoglu11}. A flurry of recent work indicates that thin films can exhibit large ferroelectric polarization\cite{wang03,jang08}, novel magnetoelectric coupling effects\cite{zhao06} and a series of strain-induced phase transitions with the promise of compelling device applications\cite{bea09,zeches09,chen10,chen11,mazumdar10,christen11}.  Currently there is particularly strong interest in the epitaxially-stabilized tetragonal-like (`$T$-like') monoclinic phase of $\mathrm{BiFeO_{3}}$ induced by large compressive strains. This Letter presents new experimental results on the nature of magnetism in essentially pure $T$-like films, and shows that the structural and antiferromagnetic phase transitions are distinct from one another, contrary to previous reports\cite{infante11,ko11}.

Epitaxial strain is known to play a major role in determining properties of thin-film multiferroic materials\cite{haeni04,choi04,schlom07}. In $\mathrm{BiFeO_{3}}$, even the slight compressive strain afforded by epitaxial growth on $\mathrm{SrTiO_{3}}$ substrates changes the structure from bulk rhombohedral to `$R$-like' monoclinic with a $\frac{c}{a}$ ratio near 1.03\cite{wang03}. The large room temperature ferroelectric moment of bulk $\mathrm{BiFeO_{3}}$ is retained, and with increasing strain the crucial out-of-plane ferroelectric moment is enhanced\cite{jang08}. Antiferromagnetic order is largely G-type\cite{bea07,ke10,ratcliff11} with a commensurability that depends on local structural considerations\cite{ratcliff11}.

When the compressive strain exceeds 4.5$\%$ a new phase appears with a $\frac{c}{a}$ ratio near 1.25\cite{bea09,zeches09}. This `$T$-like' phase can be thought of as a monoclinic distortion from a predicted meta-stable $P4mm$ tetragonal phase\cite{ederer05,ricinschi06,hatt10}. It is now understood to be part of a sequence of phases in $\mathrm{BiFeO_{3}}$: with compressive strain the structural progression is rhombohedral -- ($R$-like) monoclinic -- ($T$-like) monoclinic -- tetragonal , similar to what is observed at morphotropic phase boundaries in lead-based piezo-electric materials\cite{christen11}. As has recently been pointed out by Damodaran \textit{et al.}\cite{damodaran11}, this response to strain is very different from that of a Poisson-like deformation. In a Poisson-like elastic regime, coherent strain between a film a substrate is only maintained up to a critical thickness\cite{matthews74}. $\mathrm{BiFeO_{3}}$, in contrast, responds via polymorphism similar to the epitaxial halides\cite{farrow83} or antimony, which can be epitaxially stabilized in the ``grey tin'' allotrope rather than the stable ``white tin'' phase in the bulk\cite{farrow81}. 

While the ferroelectric transition temperature, $T_{c}$, of $R$-like films decreases with compressive strain\cite{infante10}, its value is unknown for $T$-like $\mathrm{BiFeO_{3}}$. However, such films exhibit a structural phase transition between distinct $T$-like monoclinic phases at 375 K\cite{siemons11,infante11, kreisel11,liu12}, leading to the anticipation of strong coupling effects as the various transition temperatures become comparable.

Very little is known at this point about the magnetic properties of $T$-like phase of $\mathrm{BiFeO_{3}}$, in large part because magnetometry studies of antiferromagnetic films are often inconclusive due to the strong diamagnetic response of substrate materials. First-principles calculations for films with 6$\%$ compressive strain found near-degenerate energy scales for G-type and C-type magnetic structures, where adjacent antiferromagnetic layers are stacked with spins antiparallel and parallel, respectively\cite{hatt10}. Such calculations\cite{hatt10, dieguez11} predict that structures with a large $\frac{c}{a}$ ratio, including the $T$-like phase, should exhibit ground states with C-type antiferromagnetic order\cite{dieguez11}. In contrast, a room temperature neutron diffraction measurement on one phase-pure $T$-like $\mathrm{BiFeO_{3}}$ film revealed a magnetic peak that is consistent with a G-type structure\cite{bea09}. Subsequent M\"{o}ssbauer measurements have recently been used to infer an antiferromagnetic transition at $T_{N}$ = 360 $\pm$ 20 K\cite{infante11}, within error equal to the above-mentioned structural transition. This is remarkably lower than the transitions to G-type antiferromagnetism in $R$-like films that occurs with $T_{N}$ near the bulk value of 643 K\cite{ratcliff11,infante10}.

In this Letter, we report elastic neutron scattering measurements in epitaxially-stabilized $T$-like $\mathrm{BiFeO_{3}}$ on $\mathrm{LaAlO_{3}}$ which go beyond previous studies in several important ways and provide important new information about the nature of magnetism in this material. By measuring the temperature dependence of scattering at multiple half-integer Bragg positions, we confirm the existence G-type antiferromagnetism implied by earlier measurements\cite{bea09} and provide the first estimate of N$\acute{e}$el temperature with neutrons: $T_{N}$ = 324 K. By additionally observing the T = 375 K structural transition in the same experiment, our data clearly demonstrate that the two transitions are in fact decoupled. We further present evidence for a second magnetic phase showing C-type antiferromagnetism at a lower temperature, demonstrating macroscopic phase co-existence in a $T$-like $\mathrm{BiFeO_{3}}$ film that does not contain the $R$-like polymorph. Through a comparison of scattering intensities at several magnetic Bragg peaks, it is inferred that spins in the G-type structure lie in the plane of the film.  Aided by Monte Carlo calculations, we show that the essential difference in the magnetism of $R$-like and $T$-like films can be understood using straightforward considerations of Fe-O-Fe bond distances.

An epitaxial $\mathrm{BiFeO_{3}}$ film of $300 nm$ thickness was grown on (001)-oriented $\mathrm{LaAlO_{3}}$ substrates via pulsed laser deposition. (Pseudo-cubic notations are used throughout this Letter.) Details about the growth technique and conditions are described elsewhere\cite{christen11}, as are extensive x-ray scattering investigations of the quality and structural evolution of the film investigated in the current Letter\cite{siemons11}.

The data in Ref.~\onlinecite{siemons11} (obtained on the same sample studied here) show that the film consists only of $T$-like $\mathrm{BiFeO_{3}}$ ($\frac{c}{a}$=1.23), a small ($< 2\%$) amount of a novel polymorph that reversibly arises below the T =375 K transition due to tiling considerations, a negligible ($<<0.1\%$)  of polycrystalline bismuth oxide, and \textit{no} measurable amount of the $R$-like phase. This near phase purity contrasts some previous reports of $\mathrm{BiFeO_{3}}$ films on $\mathrm{LaAlO_{3}}$, where a competing $R$-like phase occupies up to 30--60$\%$ of the sample volume\cite{zeches09,infante11}. Nonetheless, it is in agreement with numerous reports in the literature of up to 99$\%$ phase pure $T$-like $\mathrm{BiFeO_{3}}$ in samples of thickness 100nm or greater\cite{bea09,siemons11,kreisel11}.

Neutron scattering measurements were performed using the HB1a Triple-Axis Spectrometer at the High-Flux Isotope Reactor in an elastic configuration with a fixed incident energy $E_{i}$=14.7meV, pyrolytic graphite (PG) (002) monochromator and analyzer, instrument collimations 48$^{\prime}$-48$^{\prime}$-40$^{\prime}$-60$^{\prime}$, and higher order contamination removed by standard PG filters. The film was aligned in the (H H L) scattering plane. For measurements at temperatures ranging from 10 K to 375 K, the sample was mounted in a closed cycle refrigerator (CCR) in a can with thermal contact assured by He exchange gas. Additional measurements from 40 K to 435 K were done using a high-temperature CCR which lacked exchange gas and from 300 K to 500 K using a furnace. Where presented below, data from the furnace were scaled to match the data from the CCR as laid out in the supplementary material.

\begin{figure}[t]
\begin{center}
\includegraphics[width=\columnwidth]{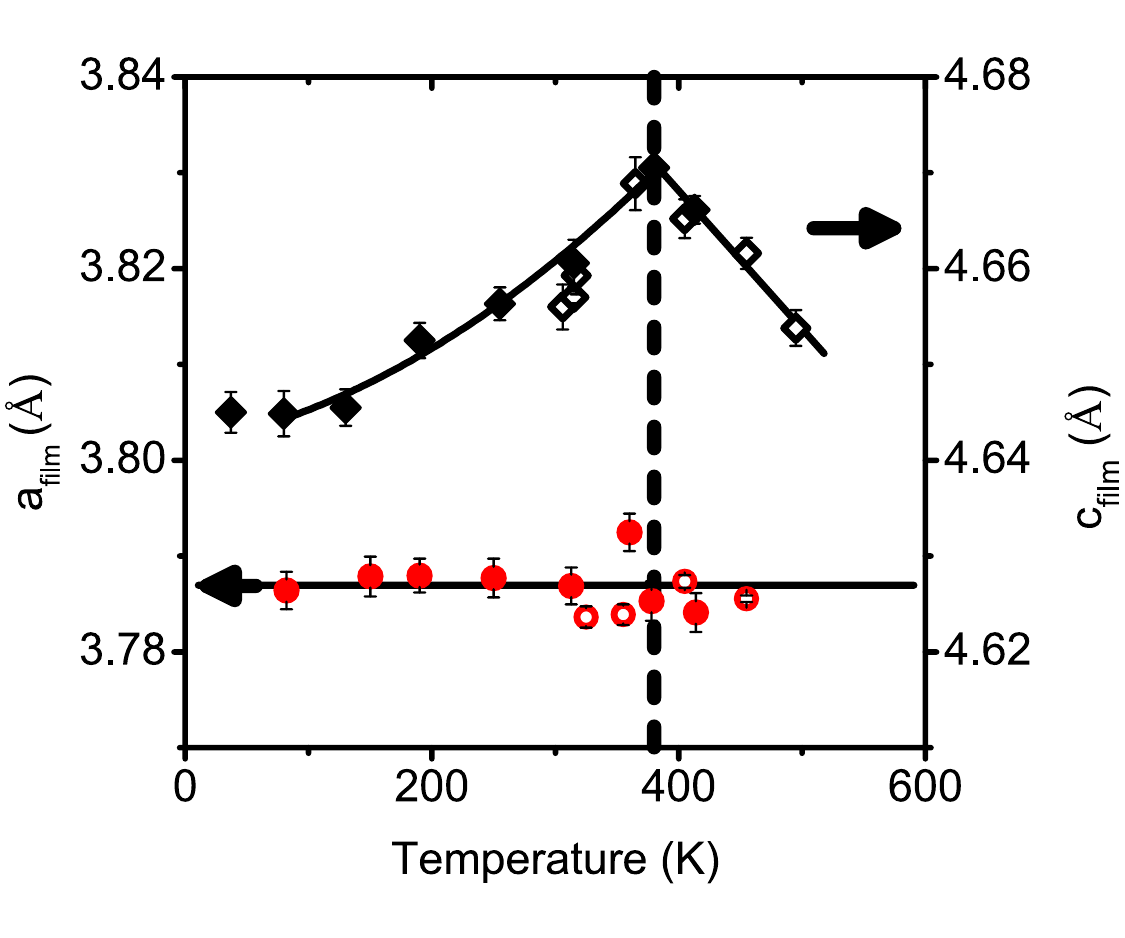}
\caption{\label{fig:structure} Temperature dependence of the out-of-plane (diamonds) and in-plane (circles) lattice parameters of the $\mathrm{BiFeO_{3}}$ film. Closed symbols represent measurements taken with the high-temperature CCR apparatus upon warming from 40 K, and open symbols represent data taken with the furnace upon cooling from 500 K. A sharp change in the temperature dependence of $c$ is seen at $T^{*}\approx$ 375 K (dashed line), without a concomitant change in $a$. This agrees with the $T$-like monoclinic to $T$-like monoclinic structural phase transition recently observed with x-ray scattering\cite{siemons11,infante11,kreisel11,liu12}.  Solid lines are guides to the eye.
}
\end{center}
\end{figure}

Neutron measurements confirmed that the $\mathrm{BiFeO_{3}}$ film had a nearly tetragonal structure with the in-plane lattice constants epitaxially matched to the substrate. The room temperature lattice  constants were $a$ = 3.79$\AA$, $c$ = 4.66$\AA$- consistent with the $\frac{c}{a}$ ratio for the $T$-like phase reported by Christen \textit{et al}\cite{christen11} and results from x-ray scattering on this sample\cite{siemons11}. Measurements of the film peak at (001) using both the high-temperature CCR and the furnace were used to track the evolution of the out-of-plane lattice parameter, $c$, with temperature. The results, shown in Fig.~\ref{fig:structure}, reveal a significant evolution of $c$ over the entire temperature range investigated, while the in-plane lattice parameter remained perfectly epitaxially matched to the substrate within resolution. A sharp cusp in the $c$-temperature curve at $T^{*} \approx 380~$K reflects the existence of a structural phase transition between different $T$-like phases of $\mathrm{BiFeO_{3}}$, as reported in recent x-ray and Raman scattering studies\cite{siemons11,infante11,kreisel11,liu12}.

\begin{figure}[t]
\begin{center}
\includegraphics[width=\columnwidth]{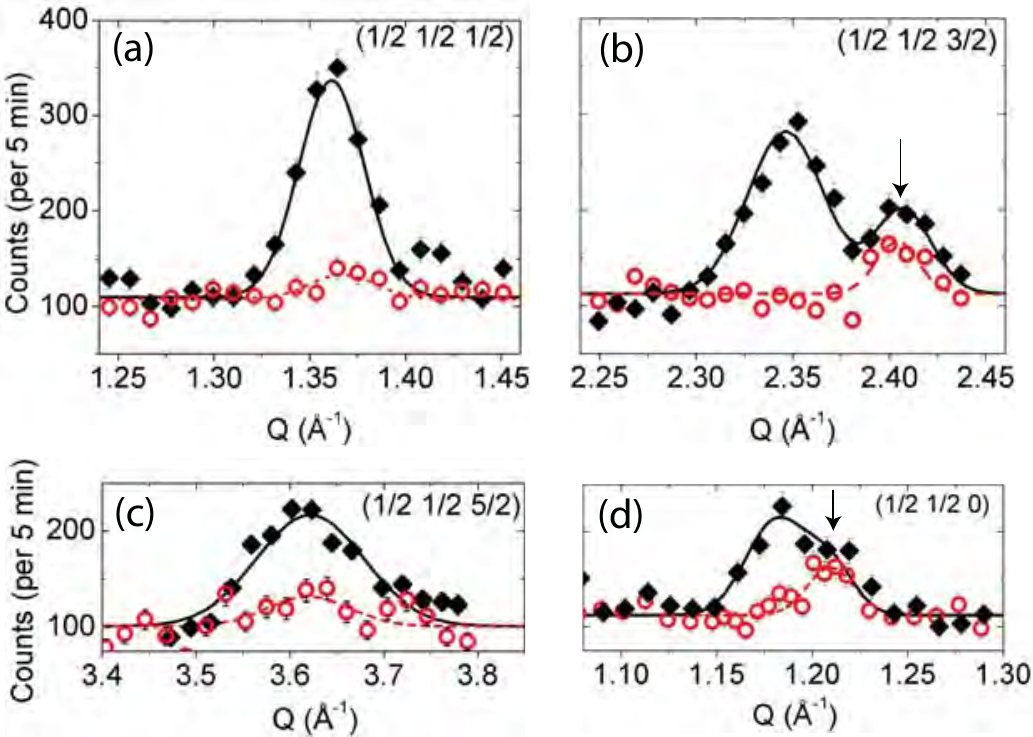}
\caption{\label{fig:raw}Elastic neutron scattering data at four half-integer film peaks at T = 370 K (open circles) and 37 K (closed diamonds) measured using the high-temperature CCR environment. Clear increased intensity at low T is seen at peaks associated with G-type antiferromagnetic order.  Shown here are: (a)($\frac{1}{2} \frac{1}{2} \frac{1}{2}$), (b)($\frac{1}{2} \frac{1}{2} \frac{3}{2}$) and (c)($\frac{1}{2} \frac{1}{2} \frac{5}{2}$). Additional intensity at low T is also observed at (d) ($\frac{1}{2} \frac{1}{2} 0$), which may indicate a small co-existing volume of C-type antiferromagnetism (see discussion in text). Arrows in panels (b) and (d) denote positions of temperature-independent background peaks. Solid lines are the results of fitting to simple Gaussians.}
\end{center}
\end{figure}

At low temperatures, neutron scattering showed superlattice peaks in the film at half-integer Bragg positions for the majority structural phase. The large $\frac{c}{a}$ ratio of the $T$-like phase allows us to clearly distinguish between signal from it and one from, for example, a possible $R$-like impurity phase that might have escaped our attention in x-ray scattering measurements. Most of the observed peaks were characteristic of G-type order: ($\frac{1}{2} \frac{1}{2} \frac{1}{2}$), ($\frac{1}{2} \frac{1}{2} \frac{3}{2}$), ($\frac{1}{2} \frac{1}{2} \frac{5}{2}$), ($\frac{3}{2} \frac{3}{2} \frac{1}{2}$) and ($\frac{3}{2} \frac{3}{2} \frac{3}{2}$). In addition, a peak was observed at the C-type wave-vector ($\frac{1}{2} \frac{1}{2} 0$). No significant signal was apparent at ($0 0 \frac{1}{2}$) or ($1 1 \frac{1}{2}$) positions, eliminating A-type antiferromagnetism. Fig.~\ref{fig:raw} shows radial scans across four peaks at temperatures above (370 K) and below (37 K) the N$\mathrm{\acute{e}}$el temperatures determined below. All panels show a significant increase of scattering intensity at low temperature. The weak peaks remaining at 370 K in panels (a) and (c) might arise from residual magnetic short range order or weak structural distortions. The temperature independent background peaks in panels (b) and (c) are powder diffraction lines of unknown origin; it has been confirmed that the background peak in panel (d) is dependent on the choice of sample environment.

Temperature dependent x-ray scattering confirmed that there is no structural transition in this sample between 300 K and 375 K\cite{siemons11}. The neutron data presented in Fig.~\ref{fig:structure} suggest no major modifications of the film structure over the entire temperature range investigated. At low temperatures, the observed intensity at ($\frac{3}{2} \frac{3}{2} \frac{3}{2}$) is slightly less than 10$\%$ of that at ($\frac{1}{2} \frac{1}{2} \frac{1}{2}$), consistent with expectations based on the magnetic form factor of Fe$^{3+}$. Moreover, the intensity ratios of the half-integer peaks are inconsistent with any of the known common structural distortions observed in cubic perovskites. The limited data set available does not permit a complete refinement of the magnetic structure, however the relative intensities of different ($\frac{1}{2} \frac{1}{2} L$) peaks imply that moments of the G-type structure are confined to the plane of the film. To within the resolution of the present measurements, no modulation of the G-type structure is visible. As mentioned above, the emergence of a peak at the ($\frac{1}{2} \frac{1}{2}$ 0) Bragg position is at odds with a simple G-type picture. We further point out that the scattering at the ($\frac{1}{2} \frac{1}{2} 0$) position is much too intense to be associated with the minority structural phase identified by x-ray scattering, which would appear an order-of-magnitude weaker in neutron scattering experiments. Since only one peak has been detected, one cannot definitively rule out a structural phase transition, albeit with no visible effect in the lattice parameters in Fig.~\ref{fig:structure}. More likely is the onset of a co-existence of two distinct magnetic ordering patterns at lower temperatures, both within the majority $T$-like structural phase. This latter possibility is consistent with the near-degeneracy of G-type and C-type structures in the \textit{ab-initio} calculations\cite{hatt10}.  A comparison of the relative intensities of ($\frac{1}{2} \frac{1}{2} \frac{1}{2}$) and ($\frac{1}{2} \frac{1}{2} 0$) peaks suggests that 15-30$\%$ of the sample volume contains C-type antiferromagnetism in this scenario, depending on the orientation of the ordered spins. As the widths of the superlattice peaks measured in low-order Brillouin zones are comparable to the width of the substrate (2 2 0) peak, it is not possible in the present experiment to infer further information about the magnetic domain sizes or correlation lengths.

\begin{figure}[t]
\begin{center}
\includegraphics[width=\columnwidth]{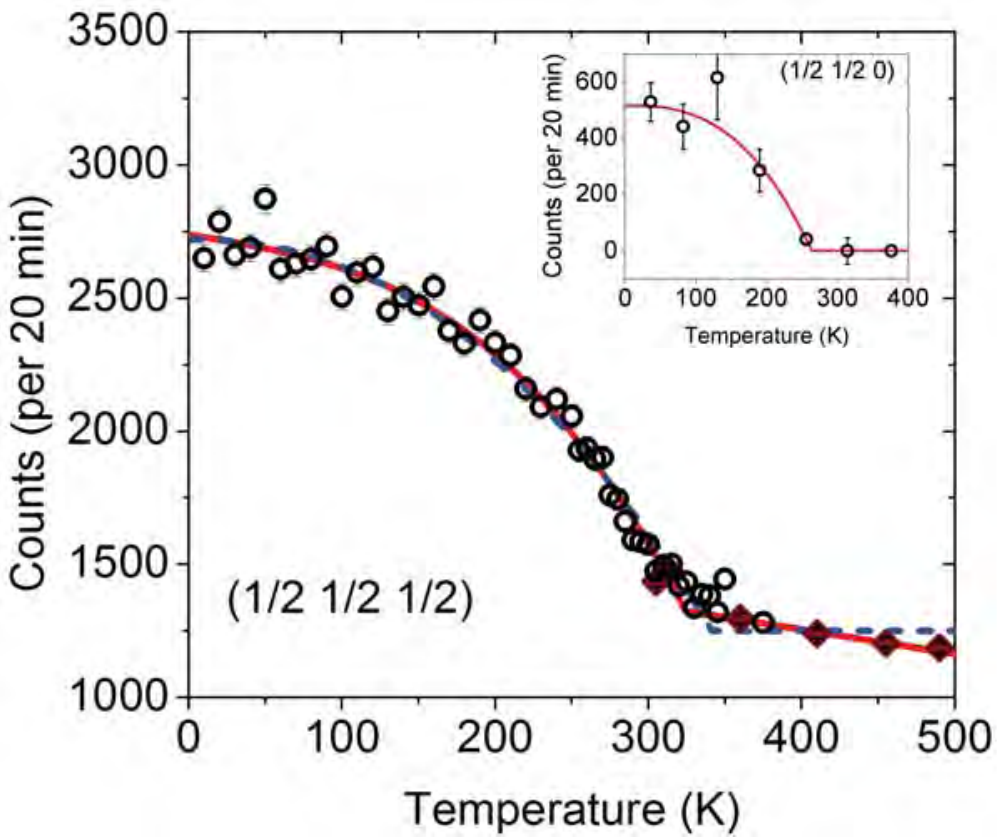}
\caption{\label{fig:OP}Plot of the neutron scattering intensity at the (1/2 1/2 1/2) peak position as a function of temperature, measured with a furnace (closed diamonds) and a low-temperature CCR (open circles) upon cooling. The data from the furnace were adjusted by scaling using the ratio of measured Bragg peak heights and adjusting the background to match the range of data overlapped with the CCR.  The solid (dashed) line is a fit assuming sloped (flat) background. Inset shows the evolution of the fitted (1/2 1/2 0) peak height over background scaled to match counting time and conditions of main panel.}
\end{center}
\end{figure}

The magnetic transition temperature of the G-type order was determined from the measured intensity of the scattering at ($\frac{1}{2} \frac{1}{2} \frac{1}{2}$), as shown in the main panel of Fig.~\ref{fig:OP}. Data from both the CCR and furnace were fit phenomenologically to a power-law temperature dependence in the range 10K$<$T$<T_N$. Particular attention was also given to the fits of background above the magnetic transition, as a small amount of temperature dependence was observed up to 500 K, possibly associated with critical correlations above a continuous phase transition. Specifically, fits were performed with either linearly sloping or flat background. The former fit, shown as a solid line in Fig.~\ref{fig:OP}, provides the superior description of the data and implies a transition temperature of $T_{N}$ = 323 $\pm$ 4 K. We believe this to be the most reliable estimate of $T_{N}$, however it should be noted that the fit with background held constant (dashed line in  Fig.~\ref{fig:OP}) yields $T_{N}$ = 341 $\pm$ 3K. The difference in these values reflects the magnitude of systematic error that might arise from different choices of background, and in fact these values may be taken as upper and lower bounds of reasonable estimates of T$_{N}$. For example, if a flat background is used and temperatures where the curve is concave-up are excluded, fits imply $T_{N}$ = 327 $\pm$ 5 K. All values represent a significant decrease from the N$\mathrm{\acute{e}}$el temperatures reported for bulk $\mathrm{BiFeO_{3}}$ (643 K)\cite{catalan09} and $R$-like films (600 - 650 K)\cite{infante10}. The magnetic transition is also clearly distinct from the structural transition identified in Fig.~\ref{fig:structure}, in contrast to some previous reports\cite{infante11,ko11}. The temperature dependence of the ($\frac{1}{2} \frac{1}{2} 0$) peak intensity, shown in the inset of Fig.~\ref{fig:OP}, suggests that the C-type order is seen only at temperatures below $T_{N,C}$ = 264 $\pm$ 12 K. (Error bars are statistical and do not include possible sources of systematic errors.) No change is seen in the intensity of the ($\frac{1}{2} \frac{1}{2} \frac{1}{2}$) scattering at $T_{N,C}$.

Simple considerations can provide insight into the different magnetic properties of the $R$-like and $T$-like phases of $\mathrm{BiFeO_{3}}$. Transition metal pseudopotential theory\cite{harrison89} predicts that the hopping matrix elements $t_{pd}$ between the $d$-orbitals of Fe and the $p$-orbitals of O scale with the Fe-O interatomic distance $r$ as $\frac{1}{r^{7/2}}$. Since perturbation theory near the atomic limit predicts that the Fe-Fe Heisenberg superexchange $J$ depends on $t_{pd}$ via $J\propto t_{pd}^{4}$, then overall $J$ varies as $J\propto r^{-14}$. Using the known structures\cite{christen11} and normalizing by the in-plane exchange constant of $R$-like $\mathrm{BiFeO_{3}}$, this criterion implies that the $R$-like superexchange is characterized by the parameters $J_{x}$=$J_{y}$=1.0 and out-of-plane $J_{z}$=0.87, while the $T$-like superexchange is characterized by $J_{x}$=1.43, $J_{y}$=2.41, and $J_{z}$=0.10. Although the precise results will be modified by additionally considering true bond angles, it is immediately apparent that the interactions in $T$-like $\mathrm{BiFeO_{3}}$ are expected to be much more two-dimensional than the interactions in the $R$-like material. Even though the in-plane values of exchange are higher for the $T$-like phase, our data imply that the net result is a suppression of magnetic order to much lower temperatures.

\begin{figure}[t]
\begin{center}
\includegraphics[width=\columnwidth]{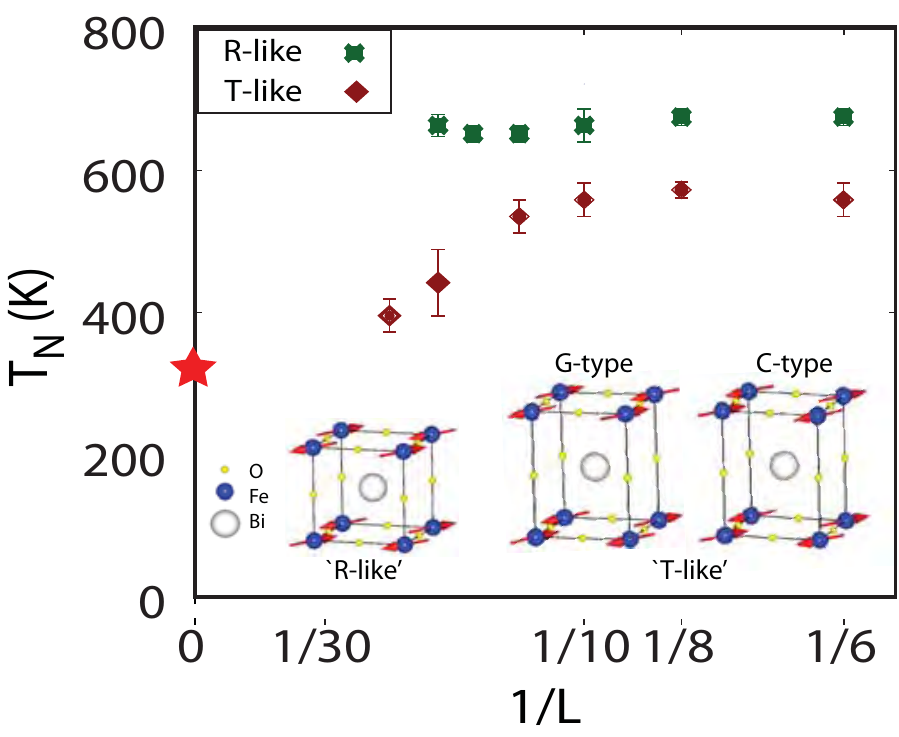}
\caption{\label{fig:MC} The convergence tendency of the ordering temperatures $T_{N}$, deduced from Monte Carlo calculations of staggered magnetization, susceptibility and heat capacity in $R$-like and $T$-like $\mathrm{BiFeO_{3}}$ films. Details about the model used in the calculations are given in the main text. The star represents the value of $T_{N}$ determined with neutrons. Also depicted are the magnetic ordering patterns observed in $R$-like and $T$-like $\mathrm{BiFeO_{3}}$, with ordered moments chosen to lie in the a-b plane.
}
\end{center}
\end{figure}

To explore this idea further, we have performed Monte Carlo simulations of the two cases utilizing an anisotropic classical Heisenberg Hamiltonian with exchange couplings in the ratios discussed above. Calculations were carried out on a set of L$\times$L$\times$L sized tetragonal lattices, with transition temperatures determined by examining the staggered magnetization, susceptibility and heat capacity. The precise magnitude of the superexchange was scaled to match the observed N$\mathrm{\acute{e}}$el temperature of the $R$-like $\mathrm{BiFeO_{3}}$ films, 643 K. Fig.~\ref{fig:MC} plots the variation of $T_{N}$ for the two cases with the Monte Carlo system size, L. The convergence to the bulk limit is much slower for the $T$-like system, as expected because of its lower effective dimensionality. However, the present Monte Carlo results are sufficient to show that the ordering for the $T$-like phase will occur at a substantially lower temperature than for the $R$-like phase, in qualitative agreement with the experimental observations.

The weak inter-planar coupling of $T$-like $\mathrm{BiFeO_{3}}$ also offers some insight into the possible co-existence of G-type and ($\frac{1}{2} \frac{1}{2}$ 0) C-type antiferromagnetism at low temperatures. Both structures (illustrated in Fig.~\ref{fig:MC}) are characterized by an antiferromagnetic arrangement of spins in the a-b plane, where the three-dimensional G-type structure is realized by antiferromagnetic stacking and C-type by ferromagnetic stacking along (00L). With very weak inter-planar interactions, one can easily suppose that small differences in local structure or impurities can lead to either state, and this is consistent with the findings of \textit{ab-initio} calculations that the energy differences are very small and may in fact favor C-type order.

In summary, the results presented here shed light on the physics of the epitaxially-stabilized $T$-like monoclinic phase of $\mathrm{BiFeO_{3}}$ which only exists in samples with high levels of compressive strain. Our film grown on $\mathrm{LaAlO_{3}}$ is nearly crystallographically phase pure, but gives evidence for phase co-existence of G-type and C-type antiferromagnetic order parameters. We estimate magnetic ordering temperature $T_{N}$ = 323 $\pm$ 4 K for the G-type order, which is distinct from the structural phase transition seen at higher temperature and more than a factor of two lower than similar order seen in $R$-like films. We attribute this large reduction mainly to a reduced dimensionality associated with increased Fe-Fe distances in the out-of-plane direction.

Research supported by the U. S. Department of Energy, Basic Energy Sciences. HMC, WS, JLZ, ED and SL were supported by the Materials Science and Engineering Division, and GJM and SEN were supported by the Scientific User Facilities Division. Experiments were performed at the High Flux Isotope Reactor and the Center for Nanophase Materials Sciences, which are sponsored by the Scientific User Facilities Division.

\clearpage
\setcounter{figure}{0}
\setcounter{equation}{0}
\setcounter{section}{0}

\renewcommand{\thesection}{\arabic{section}}

\begin{center}
\large
\textbf{SUPPORTING INFORMATION}
\end{center}

\section{Sample Characterization}

\begin{figure}[b]
\begin{center}
\includegraphics[width=\columnwidth]{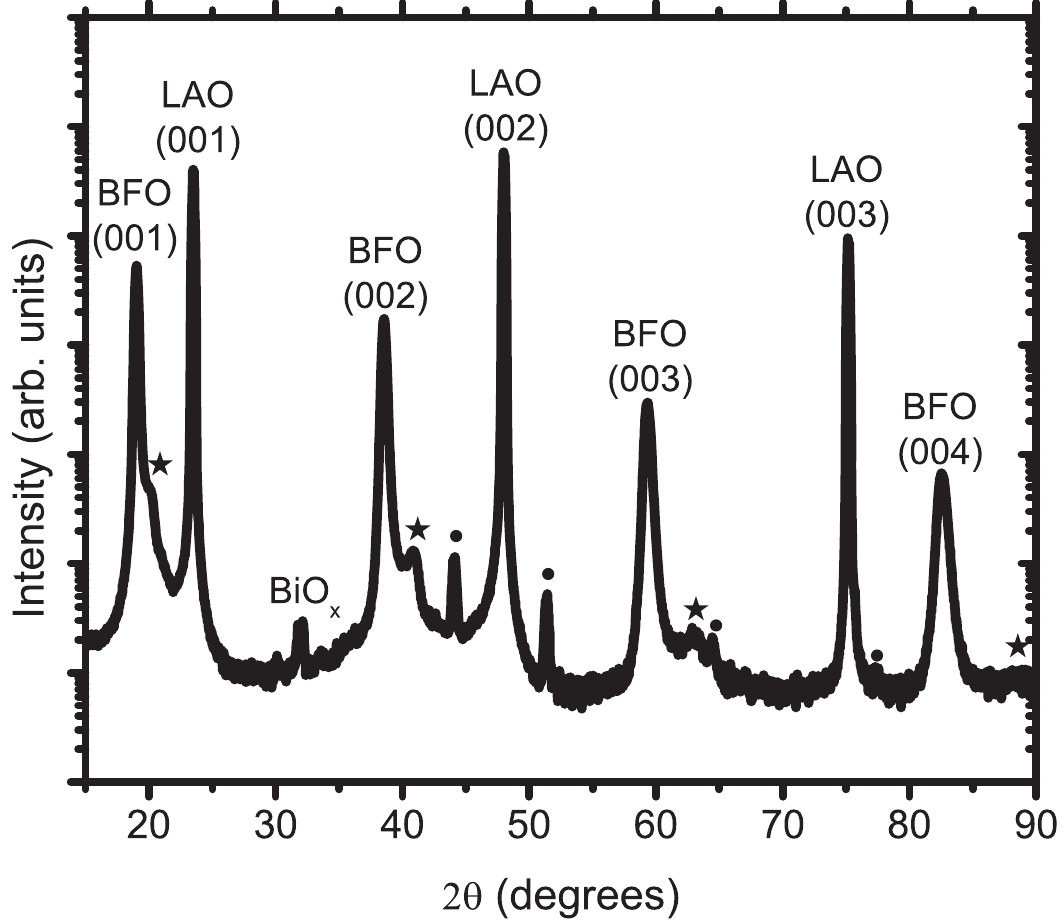}
\caption{\label{fig:xray}X-ray diffraction pattern from the sample in the current study, plotted on a logarithmic scale. Peaks arising from the film (BFO), the substrate (LAO) and the small BiO impurity have been labeled explicitly. Also present are peaks from the sample holder and a new $T$-like polymorph, labeled with black dots and stars, respectively. Data have been published previously and discussed at length in Ref.~\onlinecite{siemons11}.}\end{center}
\end{figure}

The epitaxial $\mathrm{BiFeO_{3}}$ film of the current study was grown on (001)-oriented $\mathrm{LaAlO_{3}}$ substrates via pulsed laser deposition. Thickness was determined to be 300nm ($\pm$ 20$\%$) from the spacing of x-ray thickness fringes and growth rate calculations. The sample was then characterized using a PANalytical X'Pert 4-circle thin-film x-ray diffractometer using Cu-K$\alpha$ radiation and with the sample mounted on an Anton-Paar heated stage. The resultant data have largely been published and discussed at length in Reference~\onlinecite{siemons11}. A single x-ray pattern taken at room temperature is reproduced in Fig.~\ref{fig:xray} for the purpose of demonstrating sample quality. As is easily seen, x-ray scattering data indicate that the film consists almost entirely of the $T$-like monoclinic phase of $\mathrm{BiFeO_{3}}$, with $\frac{c}{a}$=1.23. Impurity phases are reduced by nearly two orders of magnitude (note the logarithmic scale in Fig.~\ref{fig:xray}), and consist of a small ($< 2\%$) amount of a novel $T$-like polymorph of $\mathrm{BiFeO_{3}}$ that reversibly arises below the T =375 K transition from tiling considerations (see Ref.~\onlinecite{siemons11}) and a negligible ($<<0.1\%$) amount of polycrystalline bismuth oxide. There is no measurable amount of the $R$-like phase of $\mathrm{BiFeO_{3}}$, which would manifest as peaks nearer to those of $\mathrm{LaAlO_{3}}$ with $\frac{c}{a}$=1.03.

\section{Sample Environments and the Scaling of Data Sets.}

Neutron scattering data presented in the main article were obtained using three separate sample environments for controlling temperature over the range 10-500K: a high-temperature closed-cycle refrigerator (CCR) over the range 40-400 K, a high-temperature furnace over the range 300-500 K, and a low-temperature CCR with exchange gas which covered the range 10-370 K. It was demonstrated in Figure~1 of the main article that both data from the furnace and high-temperature CCR reflect the existence of the structural phase transition previously observed by x-ray scattering and can be used to independently verify the structural transition temperature of $T_{s}\sim$380 K. As the measurements of the c-axis lattice parameter were measured on warming in the high-temperature CCR and upon cooling in the furnace, this further demonstrates that there is minimal hysteresis for this transition in our sample. In neither of these high-temperature sample environments is there any signature of magnetism at or appearing immediately below the structural transition temperature.

\begin{figure}[htb]
\begin{center}
\includegraphics[width=\columnwidth]{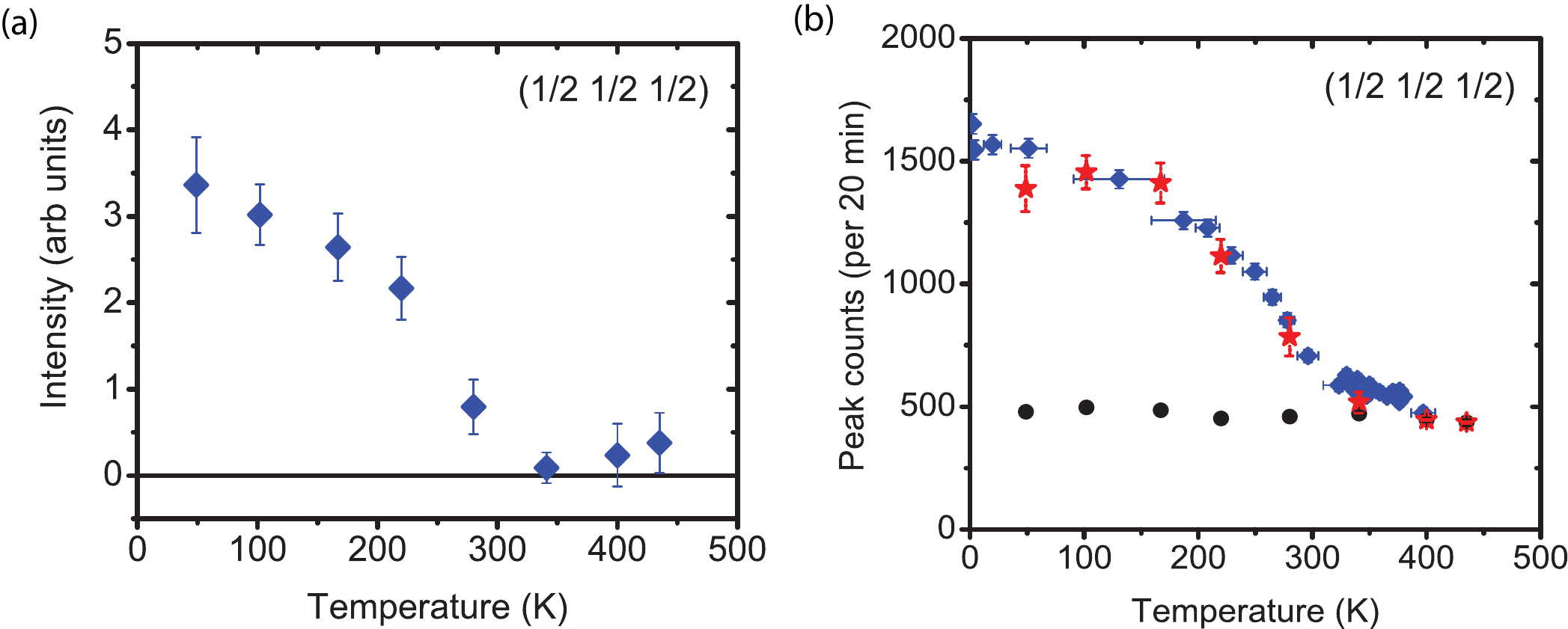}
\caption{\label{fig:OP_highT} (a) Integrated intensity of the ($\frac{1}{2} \frac{1}{2} \frac{1}{2}$) magnetic Bragg peak, as measured via radial scans taken on warming using the high-temperature CCR sample environment. (b) Peak height (red stars) and the \textbf{Q}-independent background (black circles) extracted from the same scans and plotted as a function of temperature. Also plotted is the neutron scattering intensity at the ($\frac{1}{2} \frac{1}{2} \frac{1}{2}$) taken in the same sample environment while cooling from 400 K, demonstrating the minimal amount of hysteresis observed in the current sample.}  \end{center}
\end{figure}

A measure of the N$\mathrm{\acute{e}}$el temperature for the G-type antiferromagnetically ordered phase can be determined independently using each of the three data sets and the estimates are found to be consistent with one another. One example is shown in Fig.~\ref{fig:OP_highT}(a), where plotted versus temperature is the integrated intensity of the ($\frac{1}{2} \frac{1}{2} \frac{1}{2}$) magnetic Bragg peak as measured via radial scans on warming in the high-temperature CCR. It is clear from this data that $T_N$ is above room temperature and no larger than 340 K,  consistent with the analysis presented in the main text (see Fig.~3 of the main Letter). To check for possible hysteresis effects, which would imply a lower $T_N$, we plot in Fig.~\ref{fig:OP_highT}(b) the peak height extracted from the same radial scans and compare to measurements of the scattering intensity measured while cooling in the same sample environment. The two data sets agree over an extended temperature range, implying that for this transition as well hysteresis is minimal.

For direct comparison of the different data sets, one can also scale data according to:

\begin{equation}
I^{'}(T) = \alpha (I(T)-BG_{instrument}) + \beta,
\label{eq:scale}
\end{equation}

where $\alpha$ accounts for differing effective scattering volume between separate measurements and $BG_{instrument}$ is associated with \textit{sample-independent} sources of scattering background. Fig.~\ref{fig:OP} demonstrates the result scaling data from all three sample environments to match those taken in the low-temperature CCR. Here, $BG_{instrument}$ was chosen to be the temperature independent part of the \textbf{Q}-independent background extracted from radial scans across the magnetic Bragg peak; this is a valid assumption since incoherent scattering in both $\mathrm{BiFeO_{3}}$ and $\mathrm{LaAlO_{3}}$ is minimal. The scaling factors $\alpha$ were chosen by comparing nuclear Bragg peak intensities of the film near room temperature, and $\beta$ was chosen such that magnetic data were consistent at T=370K ($T_{N} < T < T_{c}^{structural}$). Solid and dashed lines are the fits described in the main text. Of immediate note is that at high temperature all of the scattering data are consistent and imply a similar magnetic transition temperature. At lower temperatures, the scattering measured using the high-temperature CCR is systematically lower than what is measured in the low-temperature CCR. We attribute this to inefficient cooling of the sample in the high-temperature displex, expected in this sample environment at low temperatures due to the lack of exchange gas.

\begin{figure}[bth]
\begin{center}
\includegraphics[width=\columnwidth]{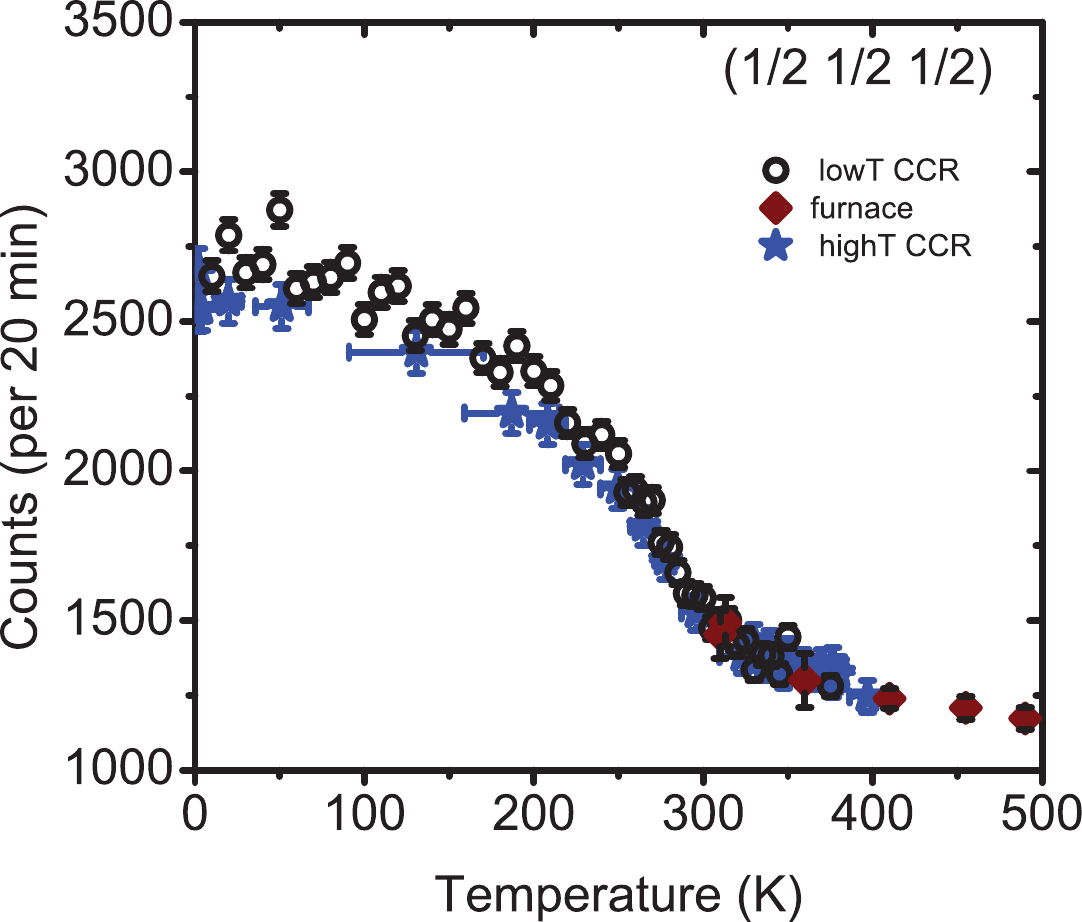}
\caption{\label{fig:OP} Plot of neutron scattering intensity at the ($\frac{1}{2} \frac{1}{2} \frac{1}{2}$) position of the $\mathrm{BiFeO_{3}}$ film as a function of temperature, including data taken in the low-T CCR (circles), the high-temperature furnace (diamonds),  all three sample environments employed in this experiment. All data were scaled to that in the low-temperature CCR, as laid out in the Supplementary text.}\end{center}
\end{figure}

\section{Determining the Neel Temperature}

When estimating $T_N$ in the main article, our analysis focussed on data obtained using the low-temperature CCR instead of the high-temperature CCR data plotted in Figure~2. This was because the low-temperature CCR data was deemed more reliable, due to both the large uncertainty in temperature during the high-T CCR measurements and the existence of helium exchange gas in the low-temperature apparatus. We additionally note the temperature-dependent scattering at the ($\frac{1}{2} \frac{1}{2} \frac{1}{2}$) position apparent in all of our data up to 500 K. This might represent short-range critical correlations, which must exist above a continuous phase transition. In modeling the temperature dependence of this short-range scattering, we have considered the cases of a linearly varying background, a temperature independent (flat) background and a flat background with the critical region excluded from fits.

In this section, we demonstrate directly that neither the choice of data set nor specific modeling of the background temperature dependence has a significant effect on conclusions of the current paper. Data from different sample environments were fit phenomenologically, assuming the follow form for magnetic scattering:

\begin{equation}
I(T) = I_0 \times \left(\frac{T_N-T}{T_{N}} \right)^{2\beta} + I_{BG},
\end{equation}

with $I_0$ and $\beta$ as fit parameters and $I_{BG}$ determined by data at temperatures T$>$380 K. Table~1 displays the results of this analysis, giving the inferred value of $T_N$ and quality of fit parameter $\chi^{2}$ for each scenario. Error bars reported here are statistical, arising from fits and not accounting for possible systematic effects. This is an especially relevant concern when considering data taken in the high-temperature CCR, as the measurements were almost certainly taken in non-equilibrium conditions due to equipment failure. One can see, however, that each fit in Table~1 implies a transition temperature in the range 322 $<$ $T_{N}$ $<$ 356 K. Even the highest estimate of $T_N$ is found to be 5 standard deviations below the known structural transition temperature at 375 K. If one excludes the less reliable high-T CCR data, this range is even more restrictive: 322$<$ $T_{N}$ $<$ 341 K. For every choice of data subset, the fits assuming sloping background were found to describe the data best. It is worth further noting that for these fits, there is very little variation in inferred $T_N$ among data sets. Our best estimate of N$\mathrm{\acute{e}}$el temperature is thus determined to be  $T_{N}$= 323$\pm$4 K.

\begin{table}[ph]
\begin{tabular}{|c|c|c|c|}
	\hline
Data set & BG function & $T_N$ (K) & $\chi^{2}$ \\
    \hline \hline
L & s & 325$\pm$3& 2.14 \\ \hline
L+F & s & 324$\pm$4& 1.90 \\ \hline
H & s & 323$\pm$7& 0.99 \\ \hline
H+F & s & 322$\pm$5& 0.87 \\ \hline
H+L+F & s & 324$\pm$3& 1.77 \\ \hline
\hline
L & fl & 342$\pm$3& 2.78 \\ \hline
L+F & fl & 341$\pm$3& 2.61 \\ \hline
H & fl & 356$\pm$4& 1.23 \\ \hline
H+F & fl & 356$\pm$4& 1.36 \\ \hline
H+L+F & fl & 344$\pm$2& 2.4 \\ \hline
\hline
L & fl-x & 327$\pm$5& 1.67 \\ \hline
L+F & fl-x & 327$\pm$5& 1.69 \\ \hline
H & fl-x & 329$\pm$12& 1.20 \\ \hline
H+F & fl-x & 329$\pm$12& 1.29 \\ \hline
H+L+F & fl-x & 325$\pm$4 & 1.70 \\ \hline

\end{tabular}
\caption{The results of fitting scattering  intensity at the ($\frac{1}{2} \frac{1}{2} \frac{1}{2}$) versus temperature, using data from different sample environments and under different assumptions about the high-temperature background. Considered are fits to the low-temperature CCR alone (L), the L-CCR plus the furnace(L+F), the high-temperature CCR alone (H), the high-temperature CCR plus the furnace (H+F) and the entirety of available data (H+L+F). Backgrounds are considered to be either sloping (s) with temperature, flat (fl) or flat with the critical temperature region excluded from the fit (fl-x). }
\end{table}


\begin{thebibliography}{22}
\expandafter\ifx\csname natexlab\endcsname\relax\def\natexlab#1{#1}\fi
\expandafter\ifx\csname bibnamefont\endcsname\relax
  \def\bibnamefont#1{#1}\fi
\expandafter\ifx\csname bibfnamefont\endcsname\relax
  \def\bibfnamefont#1{#1}\fi
\expandafter\ifx\csname citenamefont\endcsname\relax
  \def\citenamefont#1{#1}\fi
\expandafter\ifx\csname url\endcsname\relax
  \def\url#1{\texttt{#1}}\fi
\expandafter\ifx\csname urlprefix\endcsname\relax\def\urlprefix{URL }\fi
\providecommand{\bibinfo}[2]{#2}
\providecommand{\eprint}[2][]{\url{#2}}


\bibitem[{\citenamefont{Infante et~al.}(2011)\citenamefont{Infante, Juraszek,
  Fusil, Dupe, Gemeiner, Dieguez, Pailloux, Jouen, Jacquet, Geneste
  et~al.}}]{infante11}
\bibinfo{author}{\bibfnamefont{I.~C.} \bibnamefont{Infante}} \bibnamefont{et~al.},
\bibinfo{journal}{Phys. Rev. Lett.}
\textbf{\bibinfo{volume}{107}}, \bibinfo{pages}{237601}
   (\bibinfo{year}{2011}).

\bibitem[{\citenamefont{Ko et~al.}(2011)\citenamefont{Ko,Jung,He, Lee  et~al.}}]{ko11}
\bibinfo{author}{\bibfnamefont{K.~T.} \bibnamefont{Ko}} \bibnamefont{et~al.},
\bibinfo{journal}{Nature Comm.}
\textbf{\bibinfo{volume}{2}}, \bibinfo{pages}{567}
   (\bibinfo{year}{2011}).

\bibitem[{\citenamefont{Catalan and Scott}(2009)}]{catalan09}
\bibinfo{author}{\bibfnamefont{G.}~\bibnamefont{Catalan}} \bibnamefont{and}
  \bibinfo{author}{\bibfnamefont{J.~F.} \bibnamefont{Scott}},
  \bibinfo{journal}{Adv.\ Mater.} \textbf{\bibinfo{volume}{21}},
  \bibinfo{pages}{2463} (\bibinfo{year}{2009}).


\bibitem[{\citenamefont{Ramazanoglu et~al.}(2011)\citenamefont{Ramazanoglu,
  Ratcliff, Choi, Lee, Cheong, and Kiryukhin}}]{ramazanoglu11}
  See e.g. \bibinfo{author}{\bibfnamefont{M.}~\bibnamefont{Ramazanoglu}},
  \bibinfo{author}{\bibfnamefont{W.} \bibnamefont{Ratcliff II}},
  \bibinfo{author}{\bibfnamefont{Y.~J.} \bibnamefont{Choi}},
  \bibinfo{author}{\bibfnamefont{Seongsu} \bibnamefont{Lee}},
  \bibinfo{author}{\bibfnamefont{S.-W.} \bibnamefont{Cheong}},
  \bibnamefont{and} \bibinfo{author}{\bibfnamefont{V.}~\bibnamefont{Kiryukhin}},
  \bibinfo{journal}{Phys.\ Rev.\ B} \textbf{\bibinfo{volume}{83}},
  \bibinfo{pages}{174434} (\bibinfo{year}{2011}).

\bibitem[{\citenamefont{Wang et~al.}(2003)\citenamefont{Wang, Neaton, Zheng,
  Nagarajan, Ogale, Liu, Viehland, Vaithyanathan, Schlom, Waghmare
  et~al.}}]{wang03}
\bibinfo{author}{\bibfnamefont{J.}~\bibnamefont{Wang}} \bibnamefont{et~al.}, \bibinfo{journal}{Science}
  \textbf{\bibinfo{volume}{299}}, \bibinfo{pages}{1719} (\bibinfo{year}{2003}).

\bibitem[{\citenamefont{Jang et~al.}(2008)\citenamefont{Jang, Baek, Ortiz,
  Folkman, Das, Chu, Shafer, Zhang, Choudhury, Vaithyanathan et~al.}}]{jang08}
\bibinfo{author}{\bibfnamefont{H.~W.} \bibnamefont{Jang}} \bibnamefont{et~al.},
 \bibinfo{journal}{Phys.\ Rev.\ Lett.}
  \textbf{\bibinfo{volume}{101}}, \bibinfo{pages}{107602}
  (\bibinfo{year}{2008}).

\bibitem[{\citenamefont{Zhao et~al.}(2006)\citenamefont{Zhao, Scholl,
  Zavaliche, Lee, Barry, Doran, Cruz, Chu, Ederer, Spaldin et~al.}}]{zhao06}
\bibinfo{author}{\bibfnamefont{T.}~\bibnamefont{Zhao}} \bibnamefont{et~al.} \bibinfo{journal}{Nature\ Mater.}
  \textbf{\bibinfo{volume}{5}}, \bibinfo{pages}{823} (\bibinfo{year}{2006}).

\bibitem[{\citenamefont{B$\mathrm{\acute{e}}$a
  et~al.}(2009)\citenamefont{B$\mathrm{\acute{e}}$a, Dup$\mathrm{\acute{e}}$,
  Fusil, Mattana, Jacquet, Warot-Fonrose, Wilhelm, Rogalev, Petit, Cros
  et~al.}}]{bea09}
\bibinfo{author}{\bibfnamefont{H.}~\bibnamefont{B$\mathrm{\acute{e}}$a}} \bibnamefont{et~al.},
  \bibinfo{journal}{Phys.\ Rev.\ Lett.} \textbf{\bibinfo{volume}{102}},
  \bibinfo{pages}{217603} (\bibinfo{year}{2009}).

\bibitem[{\citenamefont{Zeches et~al.}(2009)\citenamefont{Zeches, Rossell,
  Zhang, Hatt, He, Yang, Kumar, Wang, Melville, Adamo et~al.}}]{zeches09}
\bibinfo{author}{\bibfnamefont{R.~J.} \bibnamefont{Zeches}} \bibnamefont{et~al.},
 \bibinfo{journal}{Science}
  \textbf{\bibinfo{volume}{326}}, \bibinfo{pages}{977} (\bibinfo{year}{2009}).

\bibitem[{\citenamefont{Chen et~al.}(2010)\citenamefont{Chen, You, Huang, Qi,
  Wang, Sritharan, and Chen}}]{chen10}
\bibinfo{author}{\bibfnamefont{Z.}~\bibnamefont{Chen}} \bibnamefont{et~al.},
  \bibinfo{journal}{Appl.\ Phys.\ Lett.} \textbf{\bibinfo{volume}{96}},
  \bibinfo{pages}{252903} (\bibinfo{year}{2010}).

\bibitem[{\citenamefont{Chen et~al.}(2011)\citenamefont{Chen, Luo, Huang, Qi,
  Yang, You, Hu, Wu, Wang, Gao et~al.}}]{chen11}
\bibinfo{author}{\bibfnamefont{Z.~H.} \bibnamefont{Chen}} \bibnamefont{et~al.},
  \bibinfo{journal}{Adv.\ Funct.\ Mater.} \textbf{\bibinfo{volume}{21}},
  \bibinfo{pages}{133} (\bibinfo{year}{2011}).

\bibitem[{\citenamefont{Mazumdar et~al.}(2010)\citenamefont{Mazumdar, Shelke,
  Iliev, Jesse, Kumar, Kalinin, Baddorf, and Gupta}}]{mazumdar10}
\bibinfo{author}{\bibfnamefont{D.}~\bibnamefont{Mazumdar}} \bibnamefont{et~al.},
  \bibinfo{journal}{Nano Lett.} \textbf{\bibinfo{volume}{10}},
  \bibinfo{pages}{2555} (\bibinfo{year}{2010}).

\bibitem[{\citenamefont{Christen et~al.}(2011)\citenamefont{Christen, Nam, Kim,
  Hatt, and Spaldin}}]{christen11}
\bibinfo{author}{\bibfnamefont{H.~M.} \bibnamefont{Christen}},
  \bibinfo{author}{\bibfnamefont{J.~H.} \bibnamefont{Nam}},
  \bibinfo{author}{\bibfnamefont{H.~S.} \bibnamefont{Kim}},
  \bibinfo{author}{\bibfnamefont{A.~J.} \bibnamefont{Hatt}}, \bibnamefont{and}
  \bibinfo{author}{\bibfnamefont{N.~A.} \bibnamefont{Spaldin}},
  \bibinfo{journal}{Phys.\ Rev.\ B} \textbf{\bibinfo{volume}{83}},
  \bibinfo{pages}{144107} (\bibinfo{year}{2011}).

\bibitem[{\citenamefont{Haeni et~al.}(2004)\citenamefont{Haeni, Irvin, Chang,
  Uecker, Reiche, Li, Choudhury, Tian, Hawley, Craigo et~al.}}]{haeni04}
\bibinfo{author}{\bibfnamefont{J.~H.} \bibnamefont{Haeni}} \bibnamefont{et~al.},
  \bibinfo{journal}{Nature}
  \textbf{\bibinfo{volume}{430}}, \bibinfo{pages}{758} (\bibinfo{year}{2004}).

\bibitem[{\citenamefont{Choi et~al.}(2004)\citenamefont{Choi, Biegalski, Li,
  Sharan, Schubert, Uecker, Reiche, Chen, Pan, Gopalan et~al.}}]{choi04}
\bibinfo{author}{\bibfnamefont{K.~J.} \bibnamefont{Choi}} \bibnamefont{et~al.},
  \bibinfo{journal}{Science}
  \textbf{\bibinfo{volume}{306}}, \bibinfo{pages}{1005} (\bibinfo{year}{2004}).

\bibitem[{\citenamefont{Schlom et~al.}(2007)\citenamefont{Schlom, Chen, Eom,
  Rabe, Streiffer, and Triscone}}]{schlom07}
\bibinfo{author}{\bibfnamefont{D.~G.} \bibnamefont{Schlom}},
  \bibinfo{author}{\bibfnamefont{L.-Q.} \bibnamefont{Chen}},
  \bibinfo{author}{\bibfnamefont{C.-B.} \bibnamefont{Eom}},
  \bibinfo{author}{\bibfnamefont{K.~M.} \bibnamefont{Rabe}},
  \bibinfo{author}{\bibfnamefont{S.~K.} \bibnamefont{Streiffer}},
  \bibnamefont{and} \bibinfo{author}{\bibfnamefont{J.-M.}
  \bibnamefont{Triscone}}, \bibinfo{journal}{Annu.\ Rev.\ Mater.\ Res.}
  \textbf{\bibinfo{volume}{37}}, \bibinfo{pages}{589} (\bibinfo{year}{2007}).

\bibitem[{\citenamefont{Bea et~al.}(2007)\citenamefont{Bea, Bibes, Petit,
  Kreisel, and Barthelemy}}]{bea07}
\bibinfo{author}{\bibfnamefont{H.} \bibnamefont{B$\acute{e}$a}},
  \bibinfo{author}{\bibfnamefont{M.} \bibnamefont{Bibes}},
  \bibinfo{author}{\bibfnamefont{S.} \bibnamefont{Petit}},
  \bibinfo{author}{\bibfnamefont{J.} \bibnamefont{Kreisel}},
  \bibnamefont{and} \bibinfo{author}{\bibfnamefont{A.}
  \bibnamefont{Barth$\acute{e}$l$\acute{e}$my}}, \bibinfo{journal}{Phil.\ Mag.\ Lett.}
  \textbf{\bibinfo{volume}{87}}, \bibinfo{pages}{165} (\bibinfo{year}{2007}).

\bibitem[{\citenamefont{Ke et~al.}(2010)\citenamefont{Ke, Zhang, Baek,
  Zarestky, Tian, and Eom}}]{ke10}
\bibinfo{author}{\bibfnamefont{X.} \bibnamefont{Ke}},
  \bibinfo{author}{\bibfnamefont{P.~P.} \bibnamefont{Zhang}},
  \bibinfo{author}{\bibfnamefont{S.~H.} \bibnamefont{Baek}},
  \bibinfo{author}{\bibfnamefont{J.} \bibnamefont{Zarestky}},
   \bibinfo{author}{\bibfnamefont{W.} \bibnamefont{Tian}},
  \bibnamefont{and} \bibinfo{author}{\bibfnamefont{C.~B.}
  \bibnamefont{Eom}}, \bibinfo{journal}{Phys.\ Rev.\ B}
  \textbf{\bibinfo{volume}{82}}, \bibinfo{pages}{134448} (\bibinfo{year}{2010}).

\bibitem[{\citenamefont{Ratcliff et~al.}(2011)\citenamefont{Ratcliff,
  Kan, Chen, Watson, Chi, Erwin, McIntyre, Capelli, and Takeuchi}}]{ratcliff11}
  \bibinfo{author}{\bibfnamefont{W.} \bibnamefont{Ratcliff II}}~\bibnamefont{et~al.},
  \bibinfo{journal}{Adv.\ Funct.\ Mater} \textbf{\bibinfo{volume}{21}},
  \bibinfo{pages}{1567} (\bibinfo{year}{2011}).

\bibitem[{\citenamefont{Ederer and Spaldin}(2005)}]{ederer05}
\bibinfo{author}{\bibfnamefont{C.}~\bibnamefont{Ederer}} \bibnamefont{and}
  \bibinfo{author}{\bibfnamefont{N.~A.} \bibnamefont{Spaldin}},
  \bibinfo{journal}{Phys.\ Rev.\ Lett.} \textbf{\bibinfo{volume}{95}},
  \bibinfo{pages}{257601} (\bibinfo{year}{2005}).

\bibitem[{\citenamefont{Ricinschi et~al.}(2006)\citenamefont{Ricinschi, Yun,
  and Okuyama}}]{ricinschi06}
\bibinfo{author}{\bibfnamefont{D.}~\bibnamefont{Ricinschi}},
  \bibinfo{author}{\bibfnamefont{K.~Y.} \bibnamefont{Yun}}, \bibnamefont{and}
  \bibinfo{author}{\bibfnamefont{M.}~\bibnamefont{Okuyama}},
  \bibinfo{journal}{J.\ Phys.:\ Condens.\ Matter}
  \textbf{\bibinfo{volume}{18}}, \bibinfo{pages}{L97} (\bibinfo{year}{2006}).

\bibitem[{\citenamefont{Hatt et~al.}(2010)\citenamefont{Hatt, Spaldin, and
  Ederer}}]{hatt10}
\bibinfo{author}{\bibfnamefont{A.~J.} \bibnamefont{Hatt}},
  \bibinfo{author}{\bibfnamefont{N.~A.} \bibnamefont{Spaldin}},
  \bibnamefont{and} \bibinfo{author}{\bibfnamefont{C.}~\bibnamefont{Ederer}},
  \bibinfo{journal}{Phys.\ Rev.\ B} \textbf{\bibinfo{volume}{81}},
  \bibinfo{pages}{054109} (\bibinfo{year}{2010}).


\bibitem[{\citenamefont{Damodaran, et~al.}(2011)\citenamefont{Damodaran, Lee, Jambunathan, MacClaren, and
  Martin}}]{damodaran11}
\bibinfo{author}{\bibfnamefont{A.~R.} \bibnamefont{Damodaran}},
  \bibinfo{author}{\bibfnamefont{S.} \bibnamefont{Lee}},
  \bibinfo{author}{\bibfnamefont{K.} \bibnamefont{Jambunathan}},
  \bibinfo{author}{\bibfnamefont{S.} \bibnamefont{MacClaren}},
  \bibnamefont{and} \bibinfo{author}{\bibfnamefont{L.~W.}~\bibnamefont{Martin}},
  \eprint{arXiv:1110.3847} (\bibinfo{year}{2011}).

\bibitem[{\citenamefont{Matthews et~al.}(2011)\citenamefont{Matthews, and Blakeslee}}]{matthews74}
\bibinfo{author}{\bibfnamefont{J.~W.}~\bibnamefont{Matthews}},
\bibnamefont{and} \bibinfo{author}{\bibfnamefont{A.~E.}~\bibnamefont{Blakeslee}},
 \bibinfo{journal}{J.\ Cryst.\ Growth}
  \textbf{\bibinfo{volume}{27}}, \bibinfo{pages}{118} (\bibinfo{year}{1974}).

\bibitem[{\citenamefont{Farrow}(1983)\citenamefont{Farrow}}]{farrow83}
\bibinfo{author}{\bibfnamefont{R.~F.~C.}~\bibnamefont{Farrow}} \bibnamefont{et~al.},
 \bibinfo{journal}{J.\ Vac.\ Sci.\ Tech.\ B}
  \textbf{\bibinfo{volume}{1}}, \bibinfo{pages}{222} (\bibinfo{year}{1983}).

\bibitem[{\citenamefont{Farrow et~al.}(1981)\citenamefont{Farrow, Robertson, Williams, Cullis,
  Jones, Young, Dennis}}]{farrow81}
\bibinfo{author}{\bibfnamefont{R.~F.~C.}~\bibnamefont{Farrow}} \bibnamefont{et~al.},
 \bibinfo{journal}{J.\ Cryst.\ Growth}
  \textbf{\bibinfo{volume}{54}}, \bibinfo{pages}{507} (\bibinfo{year}{1981}).


\bibitem[{\citenamefont{He et~al.}(2011)\citenamefont{He, Chu, Heron, Yang,
  Liang, Kuo, Lin, Yu, Liang, Zeches}}]{he11}
\bibinfo{author}{\bibfnamefont{Q.}~\bibnamefont{He}} \bibnamefont{et~al.},
 \bibinfo{journal}{Nature\ Comm.}
  \textbf{\bibinfo{volume}{2}}, \bibinfo{pages}{225} (\bibinfo{year}{2011}).

\bibitem[{\citenamefont{Infante et~al.}(2010)\citenamefont{Infante, Lisenkov,
  Dupe, Bibes, Fussil, Geneste, Petit, Courtial, Juraszek, Belliache
  et~al.}}]{infante10}
\bibinfo{author}{\bibfnamefont{I.~C.} \bibnamefont{Infante}} \bibnamefont{et~al.},
  \bibinfo{journal}{Phys.\ Rev.\ Lett.}
  \textbf{\bibinfo{volume}{105}}, \bibinfo{pages}{057601}
  (\bibinfo{year}{2010}).

\bibitem[{\citenamefont{Siemons et~al.}(2011)\citenamefont{Siemons, Biegalski,
  Nam, and Christen}}]{siemons11}
\bibinfo{author}{\bibfnamefont{W.}~\bibnamefont{Siemons}},
  \bibinfo{author}{\bibfnamefont{M.~D.} \bibnamefont{Biegalski}},
  \bibinfo{author}{\bibfnamefont{J.~H.} \bibnamefont{Nam}}, \bibnamefont{and}
  \bibinfo{author}{\bibfnamefont{H.~M.} \bibnamefont{Christen}},
  \bibinfo{journal}{Appl. Phys. Express}
  \textbf{\bibinfo{volume}{4}}, \bibinfo{pages}{095801}
   (\bibinfo{year}{2011}).

\bibitem[{\citenamefont{Kreisel et~al.}(2011)\citenamefont{Kreisel, Jadhav,
  Chaix-Pluchery, Varela, Dix, Sanchez and Fontcuberta}}]{kreisel11}
\bibinfo{author}{\bibfnamefont{J.}~\bibnamefont{Kreisel}} \bibnamefont{et~al.},
  \bibinfo{journal}{J. Phys.: Condens. Matter}
  \textbf{\bibinfo{volume}{23}}, \bibinfo{pages}{342202}
   (\bibinfo{year}{2011}).

\bibitem[{\citenamefont{Liu et~al.}(2012)\citenamefont{Liu,Liang,Liang,Chen,Yang,Peng,Wang,Chu,Chen,Lee,Chang,Lin and Chu}}]{liu12}
\bibinfo{author}{\bibfnamefont{H.-J.}~\bibnamefont{Liu}} \bibnamefont{et~al.},
  \bibinfo{journal}{Phys.~Rev.~B}
  \textbf{\bibinfo{volume}{85}}, \bibinfo{pages}{014104}
   (\bibinfo{year}{2012}).

\bibitem[{\citenamefont{Dieguez et~al.}(2011)\citenamefont{Dieguez,
  Gonzalez-Vazquez, Wojdel, and Iniguez}}]{dieguez11}
\bibinfo{author}{\bibfnamefont{O.}~\bibnamefont{Dieguez}},
  \bibinfo{author}{\bibfnamefont{O.~E.} \bibnamefont{Gonzalez-Vazquez}},
  \bibinfo{author}{\bibfnamefont{J.~C.} \bibnamefont{Wojdel}},
  \bibnamefont{and} \bibinfo{author}{\bibfnamefont{J.}~\bibnamefont{Iniguez}},
  \bibinfo{journal}{Phys.\ Rev.\ B} \textbf{\bibinfo{volume}{83}},
  \bibinfo{pages}{094105} (\bibinfo{year}{2011}).

\bibitem[{\citenamefont{Harrison}(1989)}]{harrison89}
\bibinfo{author}{\bibfnamefont{W.~A.} \bibnamefont{Harrison}},
  \emph{\bibinfo{title}{Electronic Structure and the Properties of Solids},},
  \bibinfo{chapter}{Chapter 19}(\bibinfo{publisher}{Dover, New York}, \bibinfo{year}{1989}).

\end{thebibliography}

\begin{thebibliography}{22}
\expandafter\ifx\csname natexlab\endcsname\relax\def\natexlab#1{#1}\fi
\expandafter\ifx\csname bibnamefont\endcsname\relax
  \def\bibnamefont#1{#1}\fi
\expandafter\ifx\csname bibfnamefont\endcsname\relax
  \def\bibfnamefont#1{#1}\fi
\expandafter\ifx\csname citenamefont\endcsname\relax
  \def\citenamefont#1{#1}\fi
\expandafter\ifx\csname url\endcsname\relax
  \def\url#1{\texttt{#1}}\fi
\expandafter\ifx\csname urlprefix\endcsname\relax\def\urlprefix{URL }\fi
\providecommand{\bibinfo}[2]{#2}
\providecommand{\eprint}[2][]{\url{#2}}


\bibitem[{\citenamefont{Siemons et~al.}(2011)\citenamefont{Siemons, Biegalski,
  Nam, and Christen}}]{siemons11}
\bibinfo{author}{\bibfnamefont{W.}~\bibnamefont{Siemons}},
  \bibinfo{author}{\bibfnamefont{M.~D.} \bibnamefont{Biegalski}},
  \bibinfo{author}{\bibfnamefont{J.~H.} \bibnamefont{Nam}}, \bibnamefont{and}
  \bibinfo{author}{\bibfnamefont{H.~M.} \bibnamefont{Christen}},
  \bibinfo{journal}{Appl. Phys. Express}
  \textbf{\bibinfo{volume}{4}}, \bibinfo{pages}{095801}
   (\bibinfo{year}{2011}).


\end{thebibliography}
\end{document}